\documentclass{webofc}\usepackage[varg]{txfonts}\usepackage{bm}
\usepackage{amsmath}\usepackage{slashed}\usepackage{cancel}
\woctitle{QCD@Work 2018} 
\begin{document}\title{Global Dyson--Schwinger--Bethe--Salpeter Approach\\ to Mesons with Open Flavour}\author{Thomas Hilger\inst{1,2}\fnsep\thanks{\email{thomas.hilger@uni-graz.at}}\and Mar\'ia G\'omez-Rocha\inst{3}\fnsep\thanks{\email{mariagomezrocha@gmail.com}}\and Andreas Krassnigg\inst{2}\fnsep\thanks{\email{andreas.krassnigg@uni-graz.at}}\and\\ Wolfgang Lucha\inst{1}\fnsep\thanks{\email{Wolfgang.Lucha@oeaw.ac.at}}}

\institute{Institute for High Energy Physics, Austrian Academy of Sciences, Nikolsdorfergasse 18,\\A-1050 Vienna, Austria\and Institute of Physics, University of Graz, NAWI Graz, A-8010 Graz, Austria\and European Centre for Theoretical Studies in Nuclear Physics and Related Areas, Villa Tambosi,\\38123 Villazzano (Trento), Italy}

\abstract{Exploiting an interplay of the Bethe--Salpeter equation enabling us to regard mesons as bound states of quark and antiquark and the Dyson--Schwinger equation controlling the dressed quark propagator, we amend existing studies of quarkonia by a comprehensive description of open-flavour mesons composed of all conceivable combinations of quark flavour. Employing throughout a fixed set of model parameters, we predict some basic characteristics of these mesons, \emph{i.e.}, their masses, leptonic decay constants and corresponding in-hadron condensates entering in a generalized formulation of the Gell-Mann--Oakes--Renner relation.}\maketitle

\section{Poincar\'e-Covariant Analysis of Bound States: the Physics Case}Within the framework of quantum field theories, the Bethe--Salpeter equation, underpinned by that Dyson--Schwinger equation that governs the dressed quark propagator, offers an approach to quark--antiquark bound states in Poincar\'e-covariant manner. This quark Dyson--Schwinger equation is part of an \emph{infinite\/} tower of coupled Dyson--Schwinger equations which forces us to truncate this tower somehow to a \emph{finite\/} set of coupled relations. The merits of such a covariant approach are evident: $\star$ Needless to say, quark models constitute a very convenient formalism for comprehensive investigations of hadronic states by comparatively simple technical means. $\star$ Nonperturbative approaches get rather limited or constrained by technical or computational boundary conditions. $\star$ Covariant treatments use QCD input and modelling to bridge this gap. Understandably, the first target of a covariant approach usually is the case of quarkonia, bound states of a quark and its antiquark, and hence flavourless. In order to arrive at a comprehensive picture, we complete this kind of investigations by applying a single common formalism to \emph{all\/} conceivable flavour combinations and fathom its implications for the predicted meson masses, decay constants, and in-meson condensates, by comparison with experiment or other~findings:\begin{itemize}\item[$\bigstar$]\ Clearly, covariant descriptions of open-flavour mesons have been, and will remain, limited.\item[$\bigstar$]\ Nevertheless, such covariant studies produce utmost comprehensive sets of predictions \cite{GRHKL}.\end{itemize}

\section{Dyson--Schwinger--Bethe--Salpeter Liaison}Within the Bethe--Salpeter formalism for the Poincar\'e-covariant description of bound states, a bound state, of total momentum $P,$ composed of quark and antiquark of relative momentum $p,$ is represented by this bound state's Bethe--Salpeter \emph{amplitude\/} $\Gamma(p;P)$ or Bethe--Salpeter \emph{wave function\/} $\chi(p;P),$ being related by the dressed propagators $S_{1,2}$ of the bound-state constituents:$$\chi(p;P)\equiv S_1(p+\eta\,P)\,\Gamma(p;P)\,S_2(p-(1-\eta)\,P)\ ,\qquad\eta\in[0,1]\ .$$These dressed propagators can be found as solutions of the Dyson--Schwinger equation for the quark two-point function. In rainbow truncation, this Dyson--Schwinger equation involves the quark wave-function renormalization constant, $Z_2,$ and bare mass, $m_{\rm b},$ the transverse projector$$T_{\mu\nu}(k)\equiv\delta_{\mu\nu}-\frac{k_\mu\,k_\nu}{k^2}$$in the free gluon propagator in Landau gauge, its translationally invariant integration measure, Pauli--Villars regularized at a scale $\Lambda$ and indicated by $\int^\Lambda_q,$ and an effective coupling, $k^2\,{\mathcal{G}}(k^2),$ subsuming or mimicking all effects of both full gluon propagator and full quark--gluon vertex:$$S^{-1}(p)=Z_2\,({\rm i}\,\gamma\cdot p+m_{\rm 
b})+\frac{4}{3}\,Z_2^2\!\int^\Lambda_q\!\!{\mathcal{G}}\!\left((p-q)^2\right)T_{\mu\nu}(p-q)\,\gamma_\mu\,S(q)\,\gamma_\nu\ .$$A mass renormalization $Z_m$ relates bare, $m_{\rm b},$ and at a scale $\mu$ renormalized, $m_{q}(\mu),$ quark mass:$$m_{\rm b}=Z_m\,m_q(\mu)\ .$$On an equal footing, Bethe--Salpeter amplitude or Bethe--Salpeter wave function of the bound state are governed by the homogeneous Bethe--Salpeter equation. For mesons, the latter reads, in order to preserve the axial-vector Ward--Takahashi identity of QCD, in its ladder truncation,$$\Gamma(p;P)=-\frac{4}{3}\,Z_2^2\!\int^\Lambda_q\!\!{\mathcal{G}}\!\left((p-q)^2\right)T_{\mu\nu}(p-q)\,\gamma_\mu\,\chi(q;P)\,\gamma_\nu\ .$$Expansion of $\Gamma(p;P)$ in Lorentz covariants reformulates this bound-state equation as a system of four (for all bound states of spin zero) or eight (for all bound states of nonzero spin) coupled equations. An estimate of part of the systematic uncertainties inherent to this treatment can~be acquired by adopting for the effective couplings $k^2\,{\mathcal{G}}(k^2)$ at least two different (rather popular) models \cite{EIa,EIb}. From the solution for some bound state's Bethe--Salpeter amplitude $\Gamma(p;P)$ and mass $M,$ we infer its decay constant $f$ and in-hadron condensate \cite{MRT}, \emph{i.e.}, the matrix~element of the chiral density $\bar q_1\,\gamma_5\,q_2$ between vacuum and bound state, below quoted in form of~$|\langle\bar q\,q\rangle|^{1/3}.$

\section{It's a Long Way to \xcancel{Tipperary} Bound States, It's a Long Way to Go}
\begin{figure}[hbt]\centering\includegraphics[scale=.3711,clip]{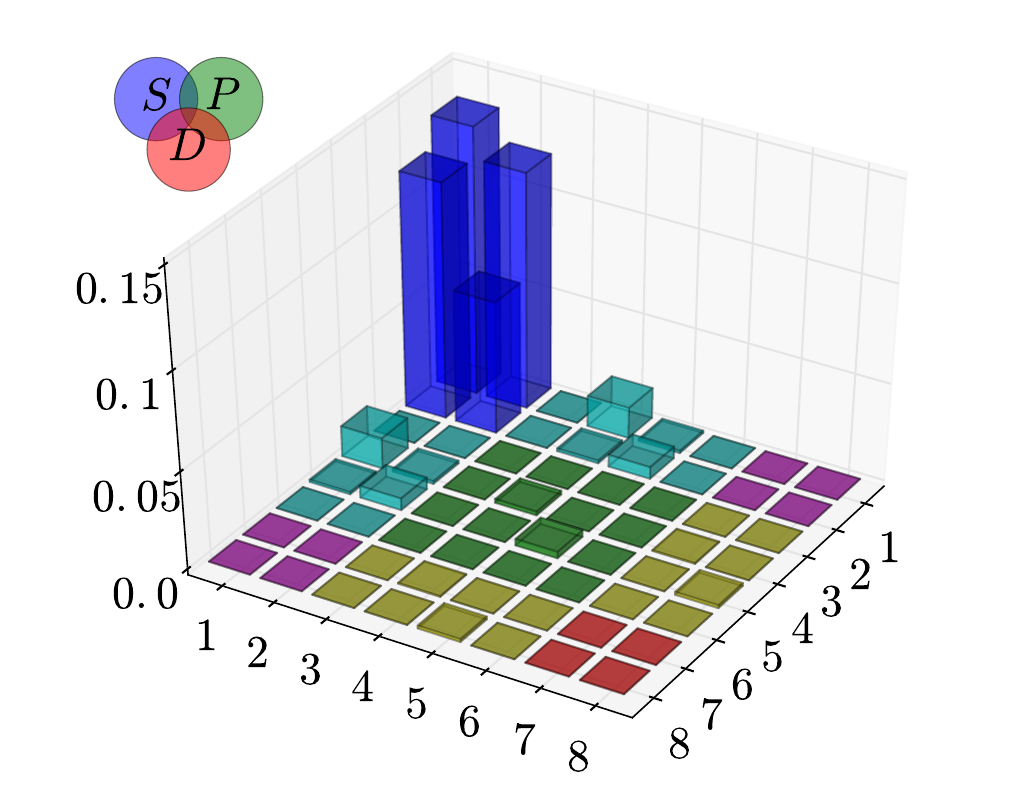} \includegraphics[scale=.3711,clip]{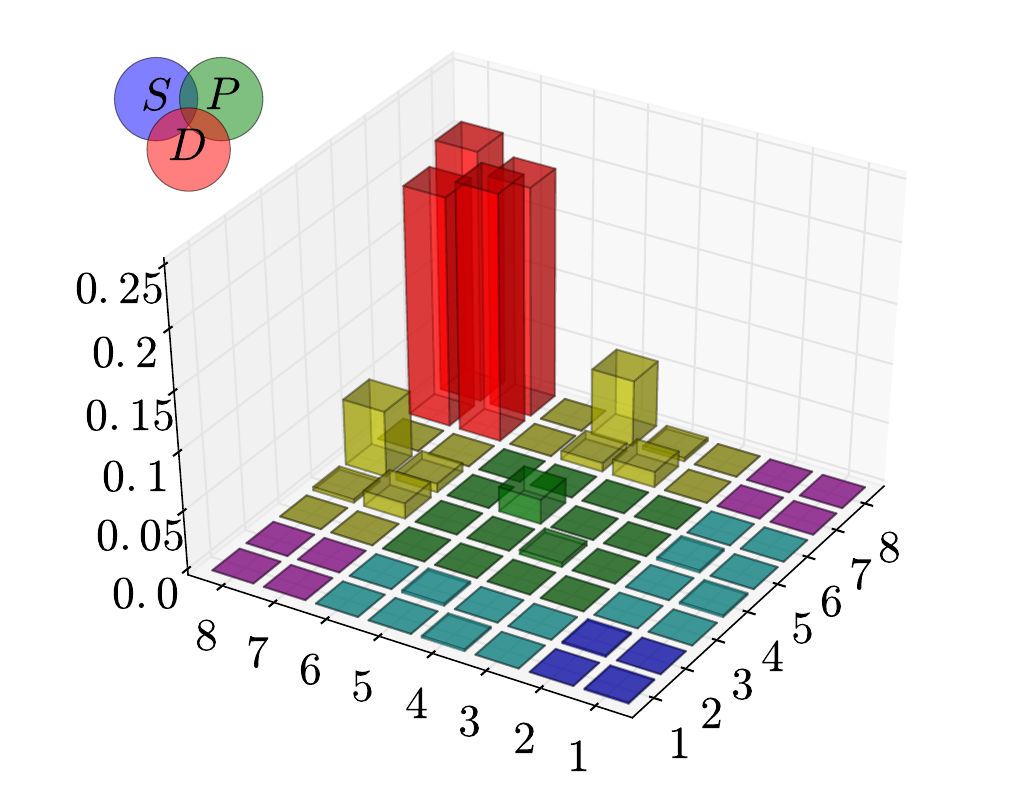}\caption{Lorentz covariants: $8\times8$ relative contributions to $\rho$ (left) or $\rho(1450)$ (right) \cite[Figs.~7 and~12]{GRHK}.}\label{l}\end{figure}

Within a covariant formalism, the classification of predicted bound states in terms of quantum numbers is not as straightforward as in a nonrelativistic treatment. Spectra of quark--antiquark mesons, with total spin $s=0,1$ and orbital angular momentum $\ell,$ can be classified, in terms of their angular momentum $J,$ parity $P=(-1)^{\ell+1},$ and charge-conjugation parity $C=(-1)^{\ell+s},$~as \begin{align*}&\mbox{ordinary:}&&J^{P\,C}\in\{0^{++},0^{-+},1^{++},1^{+-},1^{--},2^{++},2^{-+},2^{--},3^{++},3^{+-},3^{--},4^{++},4^{-+},4^{--},\ldots\}\ ,\\&\mbox{exotic:}&&J^{P\,C}\in\{0^{+-},0^{--},1^{-+},2^{+-},3^{-+},4^{+-},5^{-+},\ldots\}\ .\end{align*}This situation carries over to open-flavour mesons where one encounters \emph{quasi-exotic\/} mesons, mirroring exotic mesons of the equal-mass case \cite{HK}: the Poincar\'e-covariant setup utilized here predicts more meson states than corresponding nonrelativistic quark models. That insight can be visualized via the contributions to the norm of $\chi(p;P)$ of the different Lorentz covariants~of various values of $\ell$ attributable to them in the given meson's rest frame, as illustrated by Fig.~\ref{l}.

\section{Meson Masses, Decay Constants, and In-Hadron Condensates \cite{GRHKL}}
\begin{figure}[hbt]\centering\includegraphics[scale=.51212,clip]{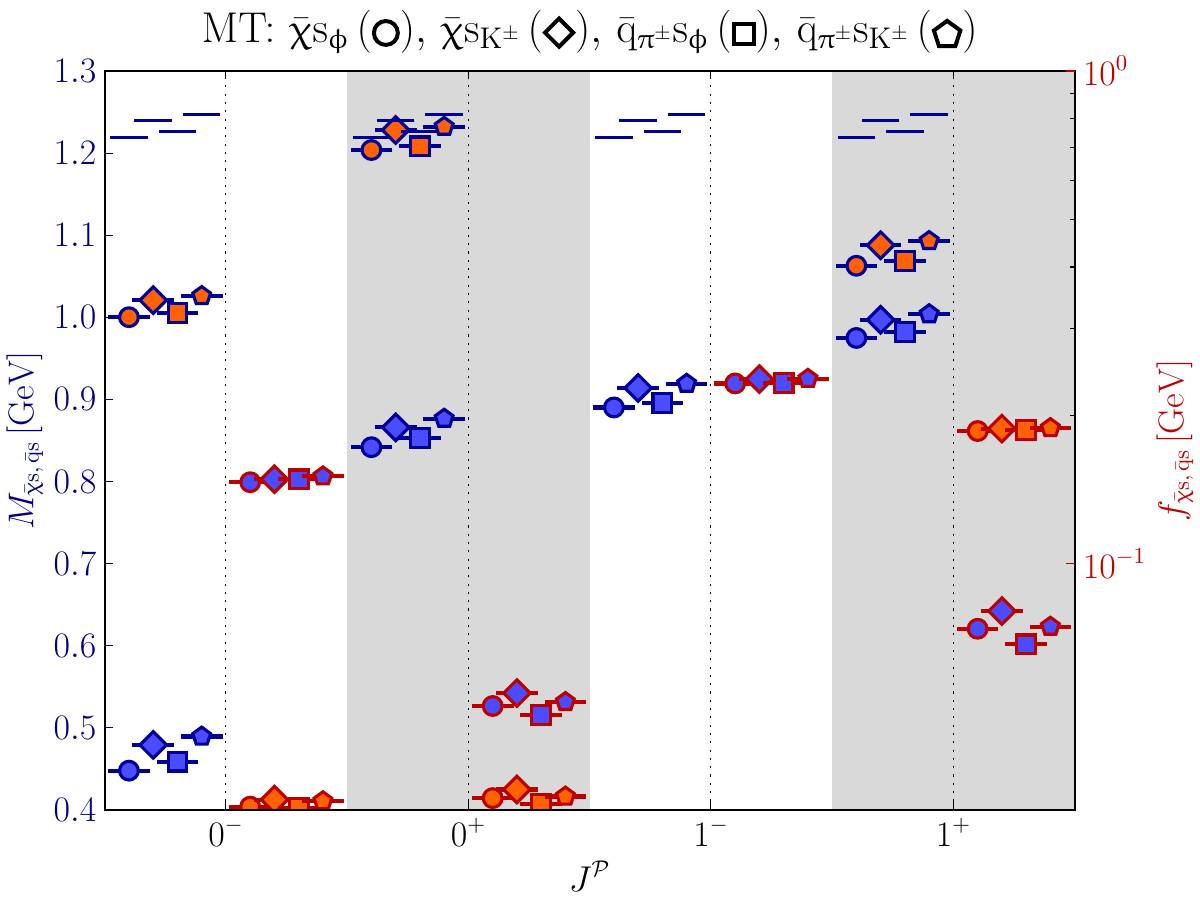}\includegraphics[scale=.22248,clip]{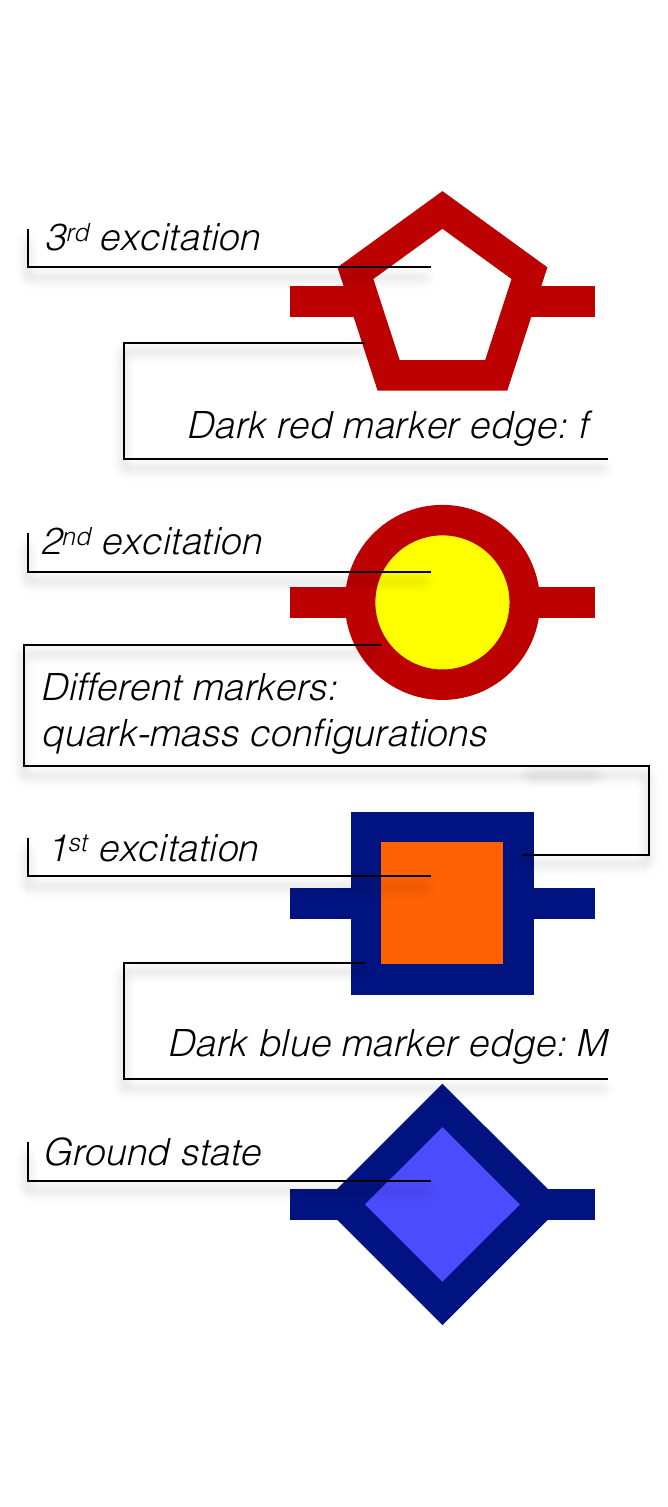}\caption{Masses and decay constants \cite[Fig.~14]{GRHKL}: mesons built up by $s$ and chiral ($\chi$) or light ($q$)~quark.}\label{qs}\end{figure}

\begin{figure}[hbt]\centering\includegraphics[scale=.51212,clip]{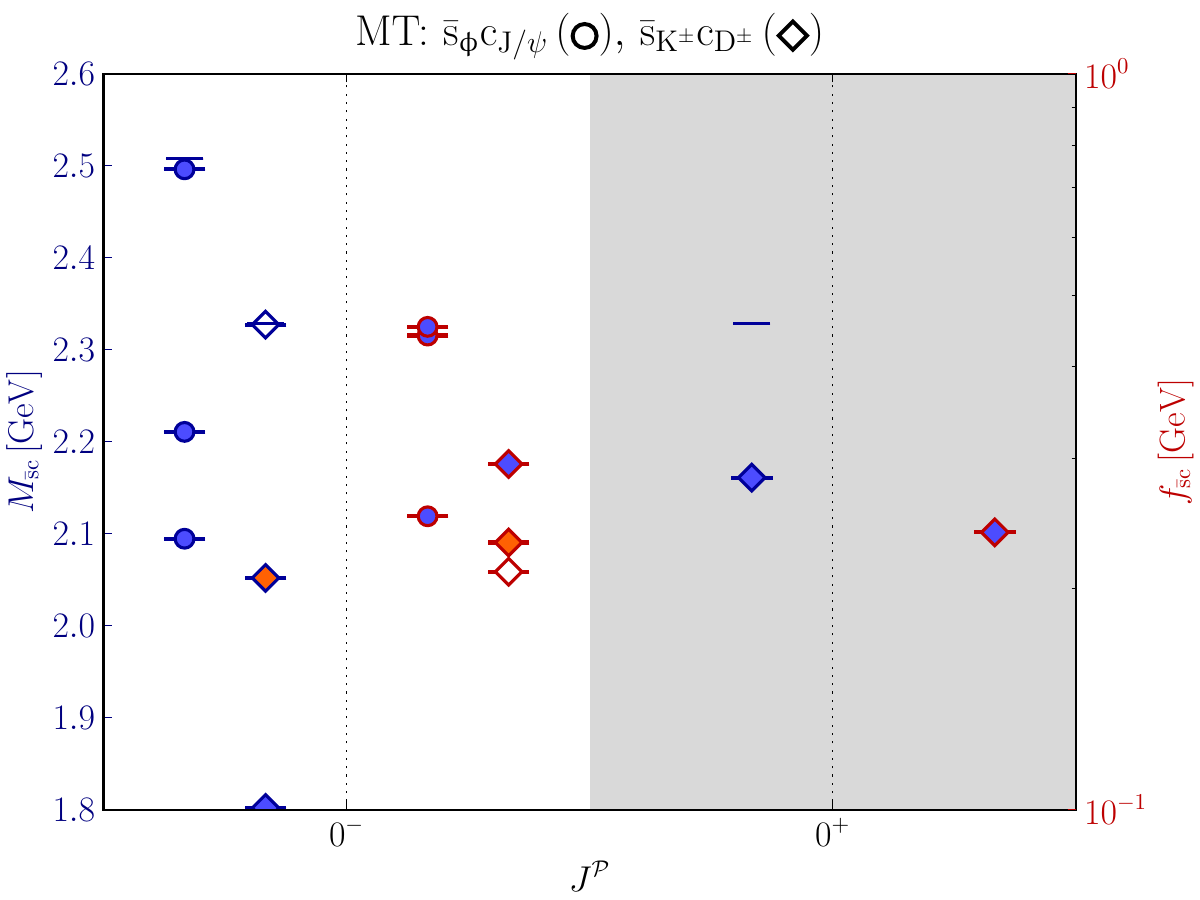}\caption{Charmed, strange mesons' masses and decay constants \cite[Fig.~17]{GRHKL} from the model of Ref.~\cite{EIa}.}\label{cs}\end{figure}

\begin{figure}[hbt]\centering\includegraphics[scale=.51212,clip]{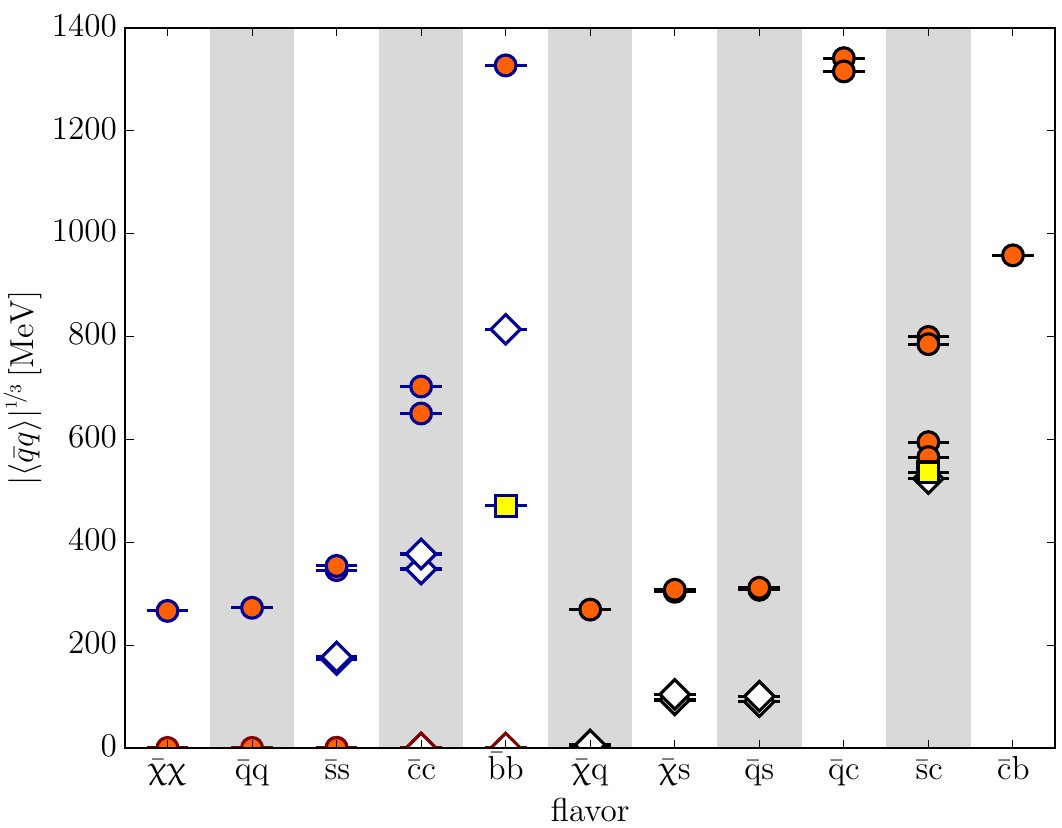}\caption{In-hadron condensates $\langle\bar q\,q\rangle$ of all mesons under study \cite[Fig.~21]{GRHKL}, from the model of Ref.~\cite{EIa}.}\label{ihc}\end{figure}

\begin{figure}[hbt]\centering\includegraphics[scale=.47864,clip]{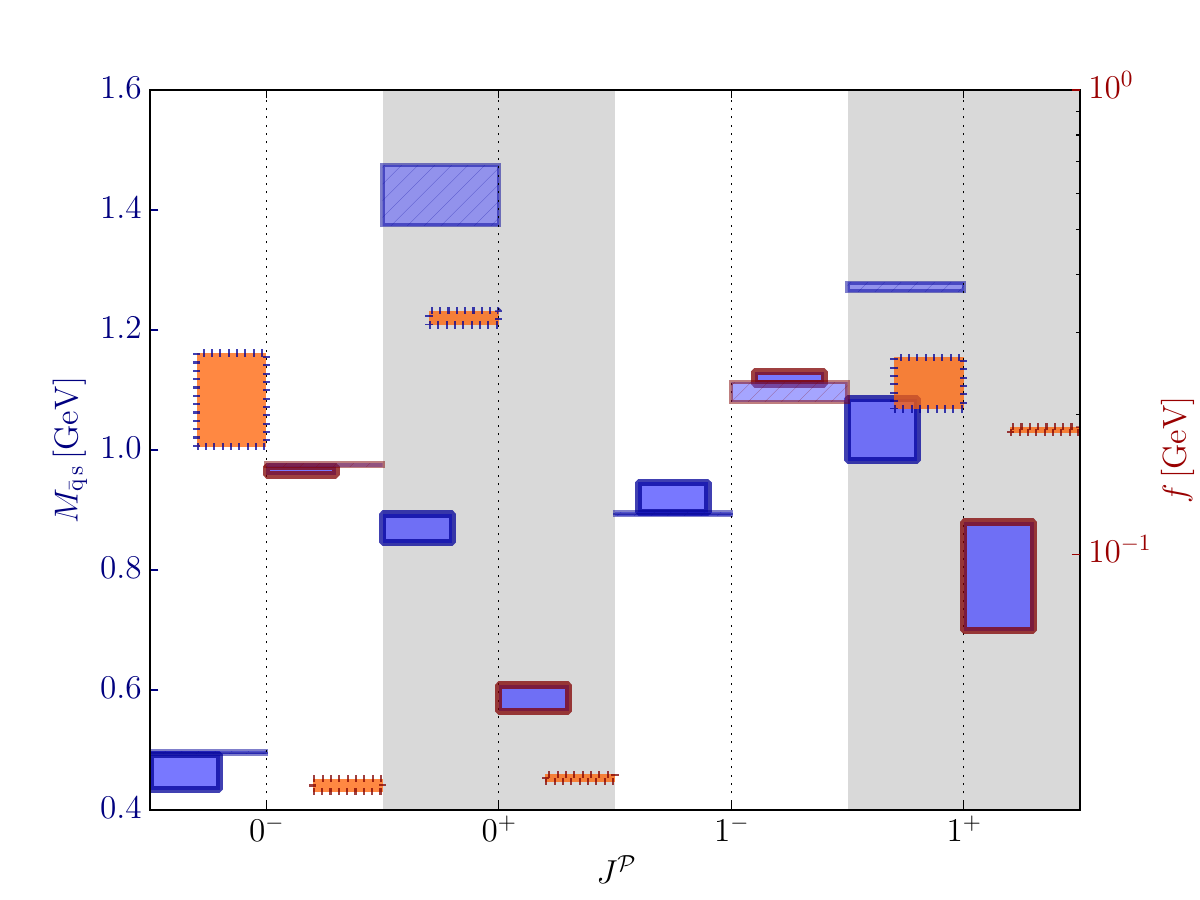}\caption{Masses and decay constants of $(\bar q\,s)$ mesons: overall result (boxes) vs.~experiment \cite[Fig.~24]{GRHKL}.}\label{qse}\end{figure}

\begin{figure}[hbt]\centering\includegraphics[scale=.47864,clip]{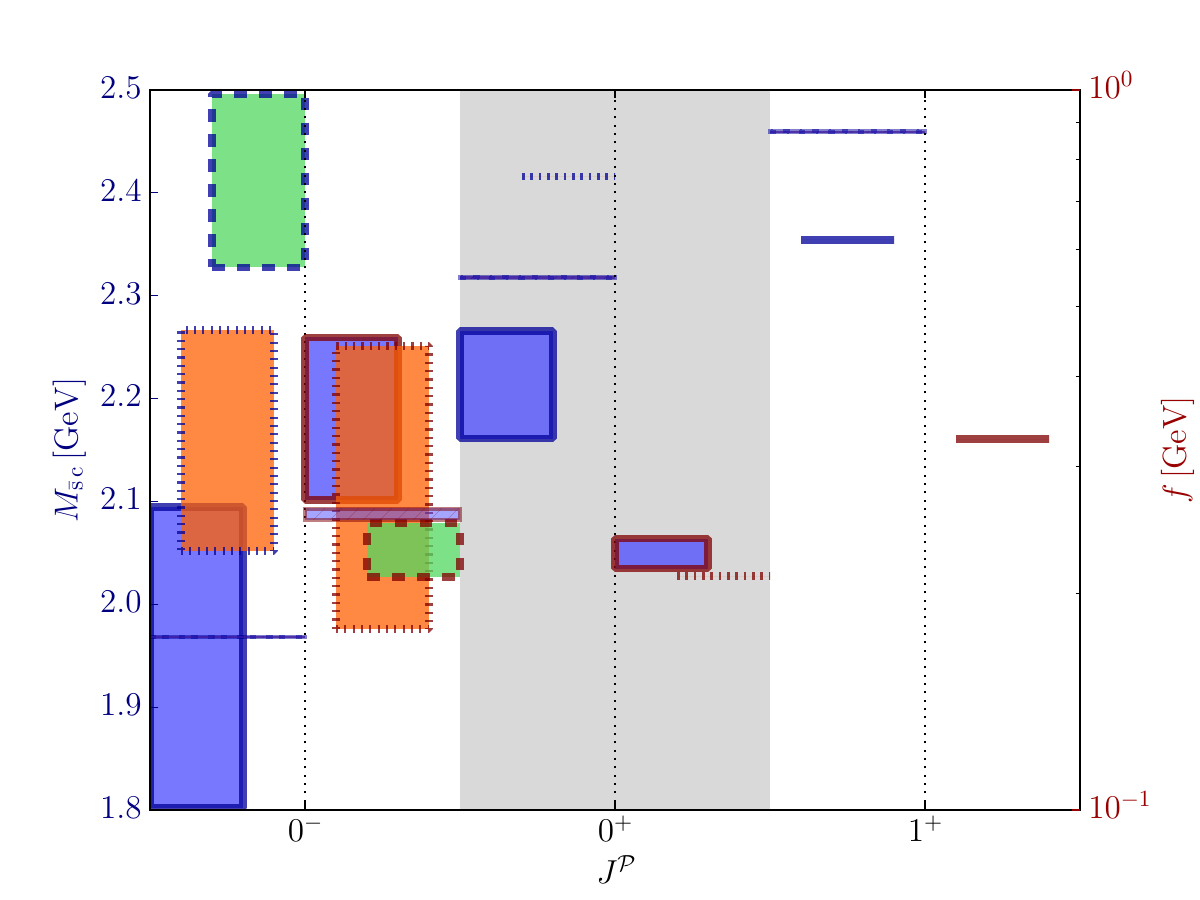}\caption{Masses and decay constants of $(\bar s\,c)$ mesons: overall result (boxes) vs.~experiment \cite[Fig.~25]{GRHKL}.}\label{cse}\end{figure}

For a given meson composed of antiquark $\bar q$ and quark $q',$ we collect our predictions \cite{GRHKL} for the masses $M_{\bar q\,q'},$ leptonic decay constants $f_{\bar q\,q'},$ and in-meson condensates $\langle\bar q\,q\rangle$ of its ground state and lowest-lying radial excitations in shared plots, separately for all flavour combinations $\bar q\,q'$ for $M_{\bar q\,q'}$ and $f_{\bar q\,q'},$ but combined to a single plot for $\langle\bar q\,q\rangle,$ as exemplified by Figs.~\ref{qs} through~\ref{ihc}. For comparison with experiment, we merge our results \cite{GRHKL} for the effective-interaction models adopted \cite{EIa, EIb} for two fits of the involved quark masses to single~predictions (\emph{cf.} Figs.~\ref{qse} and~\ref{cse}).

It goes without saying that the example outcomes shown here merely serve as a teaser: the complete sets of predictions can be found in Ref.~\cite{GRHKL}. A raw guess of the size of the~systematic uncertainties inherent to the adopted approach can be gained by variation of the model details.

\end{document}